\documentclass{article}
\usepackage{amsmath}
\usepackage{graphicx}
\usepackage{amsfonts}
\usepackage{amssymb}
\begin{document}

\title{{\Large Again on coherent states in magnetic-solenoid field}}
\author{V.G. Bagrov\thanks{Department of Physics, Tomsk State University, 634050 Tomsk, Russia; Tomsk Institute of High Current Electronics, SB RAS, 634034 Tomsk, Russia; e-mail: bagrov@phys.tsu.ru}, S.P. Gavrilov\thanks{Instituto de F\'{\i}sica, Universidade de S\~{a}o Paulo, P.O. Box 66318, 05315-970 S\~{a}o Paulo, SP, Brasil; On leave from the Department of General and Experimental Physics, Herzen State Pedagogical University of Russia, Moyka emb. 48, 191186 St. Petersburg, Russia; e-mail: gavrilovsergeyp@yahoo.com}, D.M. Gitman\thanks{Instituto de F\'{\i}sica, Universidade de S\~{a}o Paulo, P.O. Box 66318, 05315-970 S\~{a}o Paulo, SP, Brasil; e-mail: gitman@fma.if.usp.br}, and K. G\'{o}rska\thanks{Instituto de F\'{\i}sica, Universidade de S\~{a}o Paulo, P.O. Box 66318, 05315-970 S\~{a}o Paulo, SP, Brasil; On leave from H. Niewodnicza\'{n}ski Institute of Nuclear Physics, Polish Academy of Sciences, ul. Eljasza-Radzikowskiego 152, 31342 Krak\'{o}w, Poland; e-mail: kasia\_gorska@o2.pl}}

\maketitle

\begin{abstract}
\ \\
This article completes our study of coherent states in the so-called magnetic-solenoid field (a colinear combination of a constant uniform magnetic
field and Aharonov-Bohm solenoid field) presented in JPA 2010 and 2011. Here
we succeeded to prove nontrivial completeness relations for non-relativistic
and relativistic coherent states in such a field. In addition, we solve here
the relevant Stieltjes moment problem and present a comparative analysis of
our coherent states and the well-known in the case of pure uniform magnetic
field Malkin-Man'ko coherent states.
\end{abstract}

\maketitle


\section{Introduction}

A splitting of Landau levels in a superposition of the Aharonov-Bohm (AB) field and a parallel uniform magnetic field gives an example of the AB effect for bound states. In what follows, we call such a superposition the magnetic-solenoid field (MSF), more precisely MSF is a collinear combination of a constant uniform magnetic field of strength $B$ and the AB field, i.e., the field of an infinitely long and infinitesimally thin solenoid with a finite constant magnetic flux $\Phi$. Setting the $z$-axis along the AB solenoid, the MSF strength takes the form $\mathbf{B} = \left(0, 0, B_{z}\right)$, where
\begin{equation}\label{eq1}
B_{z} = B + \Phi\, \delta(x)\, \delta(y) = B + \frac{\Phi}{\pi r} \, \delta
(r),\quad B = \mathrm{const}, \quad \Phi = \mathrm{const}.
\end{equation}
We use the following electromagnetic potentials\footnote{ We accept the following notations for four- and three-vectors: $a = \left(a^{\mu}, \mu = 0, i \right) = \left(a^{0}, \mathbf{a} \right)$, $\mathbf{a} = \left(a^{i}, i = 1, 2, 3 \right) \mathbf{=} \left(a^{1} = a_{x}, a^{2} = a_{y}, a^{3} = a_{z} \right)$, $\, a_{i} = -a^{i}$, in particular, for the space-time coordinates: $x^{\mu} = (x^{0} = ct,\, x^{1} = x,\, x^{2} = y,\, x^{3} = z)$, as well as cylindrical coordinates $(r, \theta)$, in the $xy$-plane, such that $x = r \cos\theta$, $y = r \sin\theta$, and $r^{2} = x^{2} + y^{2}$. Besides, $dx = dx^{0} d\mathbf{x}\,$, $\; d\mathbf{x}\, = dx^{1} dx^{2} dx^{3}\,$, and Minkowski tensor $\eta_{\mu \nu} = \mathrm{diag}\left(  1,-1,-1,-1\right)$.} $A^{\mu}$, assigned to MSF (\ref{eq1}): $A^{0} = A^{3} = 0$, and 
\begin{equation}\label{eq2}
A_{x} = -y \left(\frac{\Phi}{2\pi r^{2}} + \frac{B}{2}\right),\quad A_{y} = x \left(\frac{\Phi}{2\pi r^{2}} + \frac{B}{2}\right),
\end{equation}
with $x = r \cos\theta$ and $y = r \sin\theta$. Henceforth, for our convenience, we will denote the flux $\Phi$ as $\Phi = \Phi_{0}(l_{0} + \mu)$, where $0 \leq \mu < 1$ and $\Phi_{0} = 2\pi c\hbar/e$ is the Dirac's fundamental unit of magnetic flux.

Solutions of the Schr\"{o}dinger equation with MSF were first studied in \cite{RRLewis83}. Solutions of relativistic wave equations (Klein-Gordon and Dirac ones) with MSF were obtained in \cite{VGBagrov01} and then used in \cite{col1} to study AB effect in cyclotron and synchrotron radiations. A profound study of these solutions and related problems can be found in \cite{HFalomir01, SPGavrilov03, SPGavrilov04-2, OLisovyy07, SPGavrilov04, PExner02} and \cite{DMGitman09}. It is important to stress that in contrast to the pure AB field case, where particles interact with the solenoid for a finite short time, moving in MSF the particles interact with solenoid permanently. This opens more possibilities to study such an interaction and correspond a number of real physical situations.

Constructing coherent states (CS) for non-relativistic and relativistic particles in the MSF is a nontrivial problem, in particular, due to the non-quadratic structure of particle Hamiltonians in this case. For the first time, CS in the MSF were constructed both in non-relativistic and relativistic cases in \cite{VGBagrov10, VGBagrov11}. However, some problems related to the constructed CS remain still open. In particular, the completeness relations for the CS were not presented. In the present article we prove these relations for non-relativistic and relativistic CS in MSF. In addition, we solve the relevant Stieltjes moment problem and present a comparative analysis of the CS in MSF and the well-known in the case of pure uniform magnetic field Malkin-Man'ko coherent states \cite{IAMalkin69}.

\section{Non-relativistic stationary states}

Let us consider a quantum behavior of a non-relativistic spinless particle with the charge $q=-e$ ($e>0$) and the mass $M$ in the MSF (see Eq.~(\ref{eq1})) in the direction perpendicular to field $B$ ($B>0$), i.e. on the $xy$-plane. As is shown in \cite{VGBagrov10} such a behavior is described by two kinds of wave functions $\Psi_{n_{1}, n_{2}}^{(j)}(t, \theta, r)$, $j = 0, 1$, namely by
\begin{equation}\label{eq3}
\Psi_{n_{1}, n_{2}}^{(j)}(t, \theta, r) = e^{-i\mathcal{E}_{n_{1}}t/\hbar}
\, \phi_{n_{1}, n_{2}}^{(j)}(\theta, \rho),\quad \rho = \frac{\gamma\, r^{2}}{2}
, \quad \gamma = \frac{eB}{c\hbar},\quad j = 0, 1,
\end{equation}
which are the eigenfunctions of two commuting to each other operators: the Hamiltonian $\hat{H}_{\perp} = \frac{1}{2M}(\hat{P}_{x}^{2} + \hat{P}_{y}^{2})$
and the angular momentum $\hat{L}_{z} = x\hat{p}_{y} - y\hat{p}_{x}$, where
$\hat{P}_{k} = \hat{p}_{k} + e A_{k}/c$, with $\hat{p}_{k} = -i \hbar \partial_{k}$, $k = x, y$, and the vector potential $\mathbf{A}$ is specified by Eq.~(\ref{eq2}). The eigenvalues of $\hat{H}_{\perp}$ and $\hat{L}_{z}$ are given by $\mathcal{E}_{n_{1}} = \hbar eB/(Mc)(n_{1}+1/2)$ and $\hbar(l-l_{0})$,
$l = 0,\pm 1,\ldots\,$, respectively. The existence of two kinds of states
$\Psi_{n_{1}, n_{2}}^{(j)}(t, \theta, r)$ is connected with the presence of AB
field ($\mu\neq 0$) and, what follows from that, the breaking translation
symmetry in $xy$-plane. The presence of a non-zero flux $\Phi$ is also visible
in two kinds of functions $\phi_{n_{1}, n_{2}}^{(j)}(\theta, \rho)$:
\begin{eqnarray}\label{eq4}
\phi_{n_{1}, n_{2}}^{(0)}(\theta,\rho) &=& \mathcal{N} e^{i(l-l_{0})\theta
}\, I_{n_{2},n_{1}}(\rho),\quad n_{1} = m, \quad n_{2} = m - l - \mu, \quad l < 0, \\[0.9pt]\nonumber
\phi_{n_{1}, n_{2}}^{(1)}(\theta, \rho) &=& \mathcal{N} e^{i(l-l_{0}) \theta - i\pi l}\, I_{n_{1}, n_{2}}(\rho), \quad n_{1} = m + l + \mu, \quad n_{2} = m, \quad l \geq 0,
\end{eqnarray}
which are orthogonal set on the $xy$-plane \cite{VGBagrov10, VGBagrov11}. $\mathcal{N} = \sqrt{\gamma/2\pi}$ is normalization constant with respect to the
inner product
\begin{equation}\label{eq5}
(f, g)_{\perp} = \int f^{\star}(\theta, \rho) g(\theta, \rho)\, dx \, dy = \frac{1}{\gamma}\, \int_{0}^{\infty} d\rho\, \int_{0}^{2\pi} d\theta\, f^{\star}(\theta, \rho)g(\theta, \rho).
\end{equation}
Here, $m = 0,1,\ldots\,$, $I_{n, m}(\rho)$ is the Laguerre function \cite{ISGradshtein07} that are related to the associated Laguerre polynomials
$L_{m}^{\alpha}(\rho)$ \cite{ISGradshtein07, APPrudnikov92-v2} as follows
\begin{equation}\label{eq6}
I_{m + \alpha, m}(\rho) = \left[\frac{m!}{\Gamma(1 + m + \alpha)}\right]
^{1/2}\, e^{-\rho/2}\, \rho^{\alpha/2}\, L_{m}^{\alpha}(\rho),\quad L_{m}^{\alpha}(\rho) = \frac{1}{m!}\, e^{\rho}\rho^{-\alpha}\, \frac{d^{m}}{d\rho^{m}}\, e^{-\rho}\rho^{m+\alpha}.
\end{equation}
The radial functions $I_{n, m}(\rho)$ were taken regularly as $r\rightarrow 0$ when $l=0$. It corresponds to a most natural self-adjoint extension of the differential symmetric operator $\hat{H}_{\bot}$. Considering a regularized case of a finite-radius solenoid one can demonstrate that the zero-radius limit yields such an extension, see \cite{SPGavrilov04}.

We know that the set of the functions $\Psi_{n_{1}, n_{2}}^{(j)}(t, \theta,r)$ is complete due to the self-adjointness of the $\hat{H}_{\bot}$. However, it is useful to show explicitly that these functions satisfy the resolution on unity (completeness relation) on $xy$-plane. For this purpose, we introduce the retarded $S^{\mathrm{ret}}(x, x^{\prime})$ Green function, which we defined as follows
\begin{equation}\label{eq7}
S^{\mathrm{ret}}(x, x^{\prime}) = \Theta(\Delta t)\, S(x,x^{\prime}),\quad S(x, x^{\prime}) = i\sum_{a} e^{-i \mathcal{E}_{n_{1}} \Delta t/\hbar}\, \phi_{n_{1}, n_{2}}^{(j)}(\theta,\rho)\, \phi_{n_{1}, n_{2}}^{\star\, (j)}(\theta^{\prime}, \rho^{\prime}), 
\end{equation}
where $\Delta t = t - t^{\prime}$ and $a$ is integers $j$, $l$, and $m$, where $l$ and $m$ are determined in Eq.~(\ref{eq4}). $\Theta(z)$ is the Heaviside step
function. Then the unity resolution for states $\phi_{n_{1}, n_{2}}^{(j)}(\varphi, \rho)$ being written with the help of $S^{\mathrm{ret}}(x, x^{\prime})$ has the form
\begin{equation}\label{eq8}
-i \left. S^{\mathrm{ret}}(x, x^{\prime}) \right\vert_{\Delta t = 0^{+}} = \delta \left(x - x^{\prime}\right)\, \delta\left(y - y^{\prime}\right).
\end{equation}
Note that $\delta \left(x - x^{\prime}\right)\, \delta\left(y - y^{\prime} \right) = \gamma\, \delta(\theta - \theta^{\prime})\, \delta(\rho - \rho^{\prime})$. It is convenient to introduce an auxiliary function $S_{l}^{(j)}(x, x^{\prime})$ by using which $S(x, x^{\prime})$ is represented as
\begin{equation}\label{eq9}
S(x, x^{\prime}) = \sum_{j = 0, 1} \sum_{l} S_{l}^{(j)}(x, x^{\prime}), \quad S_{l}^{(j)}(x, x^{\prime}) = i\, \sum_{m=0}^{\infty} e^{-i\mathcal{E}_{n_{1}}\Delta t/\hbar}\, \phi_{n_{1}, n_{2}}^{(j)}(\theta, \rho)\, \phi_{n_{1}, n_{2}}^{\star\, (j)}(\theta^{\prime}, \rho^{\prime}), 
\end{equation}
where $l<0$ for $j=0$ and $l\geq 0$ for $j=1$.

Now, by employing the states (\ref{eq4}) and formula 8.976.5 from \cite{ISGradshtein07}, we represent $S_{l}^{(j)}(x, x^{\prime})$ as:
\begin{eqnarray}\label{eq10}
S_{l}^{(j)}(x, x^{\prime}) &=& \frac{\gamma}{4\pi} \exp\left[i (l - l_{0}) \Delta \theta - i \frac{\hbar\, \gamma}{2M} (l + \mu) \Delta t\right]  \nonumber\\
&\times& \frac{\exp\left\{\frac{i}{2} (\rho + \rho^{\prime}) \cot[\hbar
\, \gamma\Delta t/(2M)] \right\}}{\sin[\hbar\, \gamma\Delta t/(2M)]}\, I_{\mp(l+\mu)} \left(\frac{\sqrt{\rho \rho^{\prime}}}{i\, \sin[\hbar
\, \gamma\Delta t/(2M)]}\right), 
\end{eqnarray}
where $\Delta\theta = \theta - \theta^{\prime}$ and $I_{\nu}(z)$ is the modified
Bessel function of the first kind. The upper sign in the index of $I_{\mp(l + \mu)}$ is related to $j=0$ and the lower is for $j=1$. The representation (\ref{eq10}) matches with the result obtained in \cite{SPGavrilov04-2}. Note that $S^{\mathrm{ret}}(x,x^{\prime})$ is the integral kernel, it can be changed from convenience considerations by changing the integration path in the complex plane of $\rho$, $\rho^{\prime}$; e.g. for $\rho = i \xi$, $\rho^{\prime} = i \xi^{\prime}$ with real positive $\xi$, $\xi^{\prime},$ we have
\begin{equation}\label{eq11}
\int d\rho\, d\rho^{\prime} S^{\mathrm{ret}}(t, t^{\prime}, \theta, \theta^{\prime}, \rho, \rho^{\prime}) f(\rho)\, g\left(\rho^{\prime} \right) = i^{2} \int d\xi\, d\xi^{\prime} S^{\mathrm{ret}}(t, t^{\prime}, \theta, \theta^{\prime}, i\xi, i\xi^{\prime}) f(i\xi)\, g(i\xi^{\prime}),
\end{equation}
where $f(\rho)$, $g\left(\rho^{\prime}\right)$ are arbitrary integrable functions. Considering the limit $\Delta t=0^{+}$ in Eq.~(\ref{eq10}), we can use the asymptotic formula (8.451.5) from \cite{ISGradshtein07} for the Bessel function. Then, going back to the initial variables, we obtain
\begin{equation}\label{eq12}
\left.S_{l}^{(j)}(x, x^{\prime})\right\vert_{\Delta t = 0^{+}} = i \frac{\gamma
}{2\pi}\, e^{i (l - l_{0}) \Delta \theta}\, \delta(\rho - \rho^{\prime}).
\end{equation}
By using the representation
\begin{equation}\label{eq13}
\frac{1}{2\pi} \sum_{l = -\infty}^{\infty} e^{i l \Delta \theta} = \delta \left(
\Delta \theta\right),
\end{equation}
we verify that relation (\ref{eq8}) holds, such that the set of the functions $\Psi_{n_{1}, n_{2}}^{(j)}(t, \theta, r)$ is really complete. Note that the distribution $S(x, x^{\prime})$ is not defined for $\Delta t = 0$, that is why the time dependent phase in Eq.~(\ref{eq7}) is important.

\section{Non-relativistic CS}

\subsection{CS in MSF}

Following the idea of \cite{VGBagrov10, VGBagrov11}, one has to introduce two kinds ($j = 0, 1$) of instantaneous CS, which are the linear combinations of the states $\phi_{n_{1}, n_{2}}^{(j)}(\theta, \rho)$ given by
Eqs.~(\ref{eq4}):
\begin{eqnarray}\label{eq14}
\mathbf{\Phi}_{z_{1}, z_{2}}^{(j)}(\theta, \rho) &=& \frac{1}{\sqrt{ \mathcal{N}_{j}(|z_{1}|^{2}, |z_{2}|^{2})}}\, \sum_{l}\mathbf{\Phi}_{z_{1}, z_{2}}^{(j), l}(\theta, \rho),\nonumber\\
\mathbf{\Phi}_{z_{1}, z_{2}}^{(j), l}(\theta, \rho) &=& \sum_{m=0}^{\infty}
\frac{z_{1}^{n_{1}}\, z_{2}^{n_{2}}}{\sqrt{\Gamma(1 + n_{1})\, \Gamma(1 + n_{2})}}\, \phi_{n_{1}, n_{2}}^{(j)}(\theta, \rho).
\end{eqnarray}
The CS are labelled by continuous complex parameters $z_{1}$ and $z_{2}$. Possible values of $n_{1}$ and $n_{2}$ depend on $m$, $l$, and $j$ according
to Eqs.~(\ref{eq4}). The normalization constants $\mathcal{N}_{j}(|z_{1}|^{2}, |z_{2}|^{2})$ can be calculated from the overlapping formula
\begin{equation}\label{eq15}
\left(\mathbf{\Phi}_{z_{1}, z_{2}}^{(j)}, \mathbf{\Phi}_{{z^{\prime}}_{1}, {z^{\prime}}_{2}}^{(j^{\prime})}\right)_{\perp} = \delta_{j, j^{\prime}}\, \frac{\mathcal{R}^{(j)}}{\sqrt{\mathcal{N}_{j}(|z_{1}|^{2}, |z_{2}|^{2}) \mathcal{N}_{j}(|z_{1}^{\prime}|^{2}, |z_{2}^{\prime}|^{2})}},
\end{equation}
where
\begin{equation}\label{eq16}
\mathcal{R}^{(0)} = Q_{1-\mu}(\sqrt{z_{1}^{\star}{z^{\prime}}_{1}},\, \sqrt
{z_{2}^{\star}{z^{\prime}}_{2}}), \quad \mathcal{R}^{(1)} = Q_{\mu}(\sqrt
{z_{2}^{\star}{z^{\prime}}_{2}},\, \sqrt{z_{1}^{\star}{z^{\prime}}_{1}}), \quad Q_{\nu}(u,v) = \sum_{l=0}^{\infty} \left(\frac{v}{u}\right)^{\nu + l} I_{\nu + l}(2uv), 
\end{equation}
for $j=j^{\prime}$ and $z_{k}={z^{\prime}}_{k}$, $k=1,2$.

Let us remark that $\mathbf{\Phi}_{z_{1} ,z_{2}}^{(j)}(\theta, \rho)$ do not represent a kind of the Gazeau-Klauder coherent states\footnote{ We recall that
GKCS are constracted on the base of a complete set of quantum states in a
specific manner, see \cite{JPGazeau99, MNovaes03}.} (GKSC).

\subsection{Completeness relations}

We are going to prove that CS (\ref{eq14}) form a complete set on $xy$-plane, that is it allows a unity resolution with the measure $d\nu_{j}(z_{1}, z_{2}) = W_{j}^{\mu}(|z_{1}|^{2}, |z_{2}|^{2})\, dz_{1}^{2}\, dz_{2}^{2}$. This statement is equivalent to the relation
\begin{eqnarray}\label{eq17}
& & \sum_{j=0,1} \left. F^{(j)}(x,x^{\prime}) \right\vert_{\Delta t=0^{+}} = -i \left.S^{\mathrm{ret}}(x, x^{\prime}) \right\vert_{\Delta t=0^{+}}, \nonumber\\
& & F^{(j)}(x, x^{\prime}) = \int\, d^{2}z_{1}\, d^{2}z_{2}\, W_{j}^{\mu}(|z_{1}
|^{2}, |z_{2}|^{2}) e^{-i \hat{H}_{\bot} \Delta t/\hbar}\, \mathbf{\Phi}_{z_{1}, z_{2}}^{(j)}(\theta, \rho)\, \mathbf{\Phi}_{z_{1}, z_{2}}^{\star \,(j)}(\theta^{\prime}, \rho^{\prime}),
\end{eqnarray}
where $W_{j}^{\mu}(|z_{1}|^{2}, |z_{2}|^{2})$ is the positive weight function and $S^{\mathrm{ret}}(x, x^{\prime})$ satisfies condition (\ref{eq8}). We consider CS defined for almost equal times. We include the time dependent phase $e^{-i \hat{H}_{\bot} \Delta t/\hbar}$ into the definition of the distribution $F^{(j)}(x, x^{\prime})$ to provide the consistency of the limit $\Delta t\rightarrow 0^{+}$ in Eq. (\ref{eq17}). To prove Eq.~(\ref{eq17}), we have to find the corresponding weight function $W_{j}^{\mu}(|z_{1}|^{2}, |z_{2}|^{2})$.

Let us check the relations
\begin{equation}\label{eq18}
\left. F^{(j)}(x, x^{\prime}) \right\vert_{\Delta t = 0^{+}} = -i\, \sum_{l}\left. S_{l}^{(j)}(x, x^{\prime}) \right\vert_{\Delta t=0^{+}},\quad j=0,1. 
\end{equation}

First, we consider the case $j=0$, for which $F^{(0)}$, after using the explicit form of $\mathbf{\Phi}_{z_{1}, z_{2}}^{(0)}(\theta, \rho)$ (see Eqs.~(\ref{eq14})), takes the form
\begin{equation}\label{eq19}
F^{(0)}(x, x^{\prime}) = \sum_{l, k}\sum_{m, n}e^{-i \hat{H}_{\bot} \Delta t/\hbar}\, \frac{\phi_{m, m-l-\mu}^{(0)}(\theta, \rho)\,\phi_{n, n-k-\mu}^{\star
\,(0)}(\theta^{\prime}, \rho^{\prime})\, G(m,n;l,k)}{\sqrt{m! n! \Gamma
(1+m-l-\mu) \Gamma(1-n-k-\mu)}}.
\end{equation}
The auxiliary function $G(m,n; l,k)$ is chosen as
\begin{eqnarray}\label{eq20}
G(m,n; l,k) &=& \int d^{2}z_{1}\, d^{2}z_{2}\, \frac{W_{0}^{\mu}(|z_{1}|^{2}, |z_{2}|^{2})}{\mathcal{N}_{0}(|z_{1}|^{2}, |z_{2}|^{2})}\, z_{1}^{m}{{z}_{1}^{\star}}^{n}\, z_{2}^{m-l-\mu}{{z}_{2}^{\star}}^{n-k-\mu
}\nonumber\\[0.01in]
&=& \frac{1}{4\pi^{2}}\, \int_{0}^{2\pi} d\varphi_{1}\, e^{i(m-n) \varphi_{1}} \int_{0}^{2\pi} d\varphi_{2}\, e^{i(m-n+k-l) \varphi_{2}} \nonumber\\[0.01in]
&\times& \int_{0}^{\infty} d|z_{1}|^{2}\, d|z_{2}|^{2}\, |z_{1}|^{m+n} |z_{2}|^{m+n-l-k-2\mu}\, \widetilde{W}_{0}^{\mu}(|z_{1}|^{2}, |z_{2}|^{2}) \nonumber\\[0.01in]
&=& \delta_{m, n}\, \delta_{l, k} \int_{0}^{\infty} du\, dv\, u^{m}\, v^{m-l-\mu
}\, \widetilde{W}_{0}^{\mu}(u, v),
\end{eqnarray}
where $z_{k} = |z_{k}| e^{i\varphi_{k}}$ ($k = 1, 2$), $u = |z_{1}|^{2}$,
$v = |z_{2}|^{2}$ and $\widetilde{W}_{0}^{\mu}(u, v) = \pi^{2}\, W_{0}^{\mu
}(u, v)/\mathcal{N}_{0}(u, v)$ which is an arbitrary positive function that
provides Eq.~(\ref{eq18}). Taking $\widetilde{W}_{0}^{\mu}(u, v) = \exp(-u-v)$
and using the representation $\Gamma(s) = \int_{0}^{\infty} x^{s-1}\, e^{-x} dx$ of the gamma function, we get
\begin{equation}\label{eq21}
G(m,n; l,k) = \delta_{m, n}\, \delta_{l, k}\, \Gamma(1 + m)\, \Gamma(1 + m - l - \mu).
\end{equation}
This function being inserted in Eq.~(\ref{eq19}) gives a correct result for (\ref{eq18}) with $j=0$.

In the same manner, one can verify the case $j=1$. Taking into account (\ref{eq8}), we see that the validity of (\ref{eq17}) is just the proof of the
completeness of CS.

We point out that the choice of $\widetilde{W}_{0}^{\mu}(u, v) = \exp(- u - v)$ in Eq.~(\ref{eq20}) produces two Stieltjes moment problems $\int_{0}^{\infty} dx\, x^{n}\, \widetilde{W}(x) = \varrho(n) = \Gamma(1+n)$, where $x$ and $n$ are
respectively taken as $u, v$ and $m, m-l-\mu$. According to Pakes's criterion \cite{AGPakes01} the appeared here Stieltjes moment problems have a unique positive solution $e^{-x}$, which leads to unambiguous, first time in the literature given, weight function $W_{j}^{\mu}(|z_{1}|^{2}, |z_{2}|^{2})$ and
at the same to unambiguous positive measure $d\nu_{j}(z_{1}, z_{1}) = W_{j}^{\mu
}(|z_{1}|^{2}, |z_{2}|^{2} )d^{2} z_{1}\, d^{2}z_{2}$.

The weight functions $W_{j}^{\mu}(u ,v)$ have the form
\begin{equation}\label{eq22}
W_{0}^{\mu}(u, v) = \pi^{-2} e^{-(u+v)}\, Q_{1-\mu}(\sqrt{u}, \sqrt{v}), \quad W_{1}^{\mu}(u, v) = \pi^{-2} e^{-(u+v)}\, Q_{\mu}(\sqrt{v},\sqrt{u}).
\end{equation}
It turns out that $W_{j}^{\mu}(u, v)$ can be expressed via special functions only for $\mu=0$ and $1/2$. The case of $\mu=0$ which corresponds to the absence of the AB filed, will be discussed in Section 3.3. In the case $\mu = 1/2$, the weight functions are
\begin{equation}\label{eq23}
W_{j}^{1/2}(u, v) = \frac{1}{2\pi^{2}}\, \left[\mathrm{erf}\left(\sqrt{u} + \sqrt{v}\right) \mp \mathrm{erf}\left(\sqrt{u} - \sqrt{v}\right)\right],
\end{equation}
where $'-'$ is for $j=0$ and $'+'$ for $j=1$. $\mathrm{erf}(z)$ is the "error
function" encountered in integrating the normal distribution \cite{ISGradshtein07}.

\subsection{Zero magnetic flux limit}

Let us study the limit $\Phi = 0$ that corresponds to the pure magnetic field without the AB solenoid.

First of all, we consider such a limit for the stationary states. All topological effects connected with the translation symmetry breaking vanish for $\mu = 0$ and, in particular, for $\Phi = 0$ ($l_{0} = 0$). As a consequence, the shift of the Landau levels is absent for $\phi_{n_{1}, n_{2}}^{(1)}(\theta, \rho)$ and it is natural to consider a superposition of $j = 1$ and
$j = 0$ states,
\begin{eqnarray}\nonumber
\phi_{m, m-l}^{(0)}(\theta, \rho) + \phi_{m+l, m}^{(1)}(\theta, \rho) = \phi_{m, l}^{L}(\theta, \rho), \quad l=0,\pm 1,\pm 2,\ldots\,.
\end{eqnarray}

Next, we study the limit of $\Phi=0$ in CS (\ref{eq14}). Thus, we expect to obtain the Malkin-Man'ko CS \cite{IAMalkin69}. To show this, we consider the following superposition of the CS:
\begin{equation}\label{eq24}
\mathbf{\Phi}_{z_{1}, z_{2}}(\theta, \rho) = \mathcal{N}_{0}^{1/2}\, \mathbf{\Phi}_{z_{1}, z_{2}}^{(0)}(\theta, \rho)\, + \,\mathcal{N}_{1}^{1/2}\, \mathbf{\Phi}_{z_{2}, z_{1}}^{(1)}(\theta, \rho).
\end{equation}
At the beginning, we note that the probability distribution of $|\mathbf{\Phi
}_{z_{1}, z_{2}}(\theta, \rho)|^{2}$ calculated with respect to the inner product $(f, g)_{\perp}$ is equal to
\begin{equation}\label{eq25}
|\mathbf{\Phi}_{z_{1}, z_{2}}(\theta, \rho)|^{2} = \mathcal{N}_{0}(|z_{1}|^{2}, |z_{2}|^{2}) + \mathcal{N}_{1}(|z_{1}|^{2}, |z_{2}|^{2}) = e^{|z_{1}|^{2} + |z_{2}|^{2}},
\end{equation}
where $\mathcal{N}_{j}(|z_{1}|^{2}, |z_{2}|^{2}) = \mathcal{R}^{(j)}$ at $z_{k} = {z^{\prime}}_{k}$ are given in Eq.~(\ref{eq16}) for $j = j^{\prime}$. To derive Eq.~(\ref{eq25}), we employ the formula 5.8.3.2 from \cite{APPrudnikov92-v2} and the fact that $\mathbf{\Phi}_{z_{1}, z_{2}}^{(j)}(\theta, \rho)$ are orthogonal for different $j$. The density
$|\mathbf{\Phi}_{z_{1}, z_{2}}(\theta, \rho)|^{2}$ is equal to the normalization
constant of Malkin-Man'ko CS, see Eq.~(41) in \cite{IAMalkin69}. Then,
substituting $\mathbf{\Phi}_{z_{1}, z_{2}}^{(j)}(\theta,\rho)$ into
Eq.~(\ref{eq25}), we obtain
\begin{equation}\label{eq26}
\mathbf{\Phi}_{z_{1}, z_{2}}(\theta, \rho) = \sum_{l, m} \frac{z_{1}^{m} z_{2}^{m+|l|}}{\sqrt{m! (m+|l|)!}}\, \left[\phi_{m, m-l}^{(0)}(\theta, \rho) + \phi_{m+l, m}^{(1)}(\theta, \rho)\right] = \sum_{r_{1}, r_{2}=0}^{\infty} \frac{z_{1}^{r_{1}}\, z_{2}^{r_{2}}}{\sqrt{r_{1}!\, r_{2}!}}\, \phi_{m, l}^{L}(\theta, \rho),
\end{equation}
where $r_{1} = m$, $r_{2} = m+|l|$. Comparing Eq.~(\ref{eq26}) with Eq.~(41) from \cite{IAMalkin69}, we see that $\mathbf{\Phi}_{z_{1}, z_{2}}(\theta, \rho)$ are just Malkin-Man'ko CS.

Now, let us consider the weight function $W_{j}^{\mu}(u, v)$ for $\mu = 0$. In
the limit under consideration, we have
\begin{equation}\label{eq27}
W_{0}^{0}(u, v) + W_{1}^{0}(u, v) = \pi^{-2}, 
\end{equation}
where the formula 5.8.3.2 from \cite{APPrudnikov92-v2} was used. Eq.~(\ref{eq27}) is the weight function $W^{0}(u, v)$ for the Malkin-Man'ko CS.

\section{Relativistic stationary states}

Note that relativistic spinless CS are reduced to the non-relativistic case. That is why in the relativistic case, only the CS of spinning particles are in a sense nontrivial. In spite of the fact that the algebra of the Dirac $\gamma$-matrices and the spin description in $(2+1)$-dim and in $(3+1)$-dim are different, considering $(3+1)$-dim case, we can use technical results obtained for $(2+1)$-dim. That is why in the beginning, we consider spinning case in $(2+1)$-dim.

The behavior of an electron in MSF in $(2+1)$-dim are described by wave functions that obey the Dirac equation with such a field, see \cite{SPGavrilov03}. These wave functions for given 'polarizations' $\xi = \pm 1$ (related to one of two nonequivalent representation for $\gamma$- matrices)
and particle/antiparticle energy $c p_{0} = \pm \mathcal{E}_{\pm}$ have the form
\begin{equation}\label{eq28}
\Psi = e^{-i(c p_{0} t)/\hbar}\, \psi_{p_{0}}^{(\xi)}(x^{1}, x^{2}).
\end{equation}
In contrast to $(3+1)$-dim case, particles and antiparticles in $(2+1)$-dim have only one spin polarization states. Choosing $\xi = + 1$, we deal with 'spin-up' particles, and choosing $\xi = - 1$ with 'spin-down' particles. One can see that $\psi_{p_{0}}^{(-1)}(x^{1}, x^{2}) = \sigma^{2}\, \psi_{-p_{0}}^{(1)}(x^{1}, x^{2})$, where $\sigma^{2}$ is a Pauli matrix. That is why we consider here only the case $\xi=1$. The 'spin-up' particle ($'+'$) and antiparticle ($'-'$) states are denoted as $\psi_{p_{0}}^{(1)} = \psi_{\pm, n_{1}, n_{2}}^{(j)}$. The functions $\psi_{\pm, n_{1}, n_{2}}^{(j)}$ are common eigenfunctions of the total angular momentum operator $\hat{J} = -i \hbar \partial_{\theta} + \hbar\sigma^{3}/2$ and of the Hamiltonian $\hat{H}^{\vartheta} = c(\mbox{\boldmath$\sigma$\unboldmath} \mathbf{\hat{P}}_{\perp} + Mc\sigma^{3})$. The eigenvalues are equal to $\hbar(l - l_{0} - 1/2)$ and $\pm \mathcal{E}_{\pm}$, respectively. $\hat{H}^{\vartheta}$ represent a one-parameter family of self-adjoint Hamiltonians (self-adjoint extensions) that are determined by the corresponding boundary conditions. We consider only two special cases: $\vartheta = \mathrm{sign}\Phi = \pm 1$. They correspond to a most natural self-adjoint extensions $\hat{H}^{\vartheta}$. Considering a regularized case of a finite-radius solenoid, one can demonstrate that the zero-radius limit yields such extensions, see \cite{SPGavrilov03}. The
functions $\psi_{\pm,n_{1}, n_{2}}^{(j)}$ can be represented as \cite{VGBagrov11}
\begin{eqnarray}\label{eq29}
\psi_{\pm, n_{1}, n_{2}}^{(j)}(\theta, \rho) &=& \mathcal{M}_{j, \pm, n_{1}, n_{2}} \left\{\sigma^{3} \left[\pm \hat{\Pi}_{0} \left(M\right) - \mbox{\boldmath$\sigma$\unboldmath} \mathbf{\hat{P}}_{\perp} \right] + M c\right\} u_{n_{1}, n_{2}, \pm 1}^{(j)}(\theta, \rho), \nonumber\\[0.01in]
\hat{\Pi}_{0} \left(M\right) &=& \hat{\Pi}_{0}^{2} \frac{2}{\sqrt{\pi}} \int_{0}^{\infty} e^{-\hat{\Pi}_{0}^{2} \tau^{2}} d\tau, \quad \hat{\Pi}_{0}^{2} 
= M^{2} c^{2} + \left(\mbox{\boldmath$\sigma$\unboldmath} \mathbf{\hat{P}}_{\perp} \right)^{2}, \nonumber\\
u_{n_{1}, n_{2}, \sigma}^{(j)}(\theta, \rho) &=& \phi_{n_{1}, n_{2}, \sigma}^{(j)}(\theta, \rho) v_{\sigma}, \quad v_{1} = \binom{1}{0}, \quad v_{-1} = \binom{0}{1}, 
\end{eqnarray}
where $\hat{\mathbf{P}}_{\perp} = (\hat{P}_{x}, \hat{P}_{y})$ and $\mathcal{M}_{j, \pm, n_{1}, n_{2}}$ are normalization factors with respect to the inner product
\begin{equation}\label{eq30}
\left(\psi,\, \psi^{\prime}\right)_{D} = \gamma^{-1} \int_{0}^{\infty} d\rho \int_{0}^{2\pi} d\theta\, \psi^{\dag}(\theta, \rho)\, \psi^{\prime}(\theta, \rho).
\end{equation}
For particles the energy spectrum is $\mathcal{E}_{+} = [(Mc^{2})^{2} + \mathcal{E}_{\perp\, (+1)}^{2}]^{1/2}$; for antiparticles the energy spectrum
is $\mathcal{E}_{-} = [(Mc^{2})^{2} + \mathcal{E}_{\perp\, (-1)}^{2}]^{1/2}$. The energy $\mathcal{E}_{\perp\, (\sigma)}^{2}$ is given by $\mathcal{E}_{\perp\, (\sigma)}^{2} = 2\hbar c e B[n_{1} + (1+\sigma)/2]$. The functions $\phi_{n_{1}, n_{2}, \sigma}^{(j)}$ have the form \cite{VGBagrov11}:
\begin{eqnarray}\label{eq31}
\phi_{n_{1}, n_{2},\sigma}^{(0)}(\theta, \rho) &=& e^{i \left(l_{\sigma} - l_{0}\right) \theta}\, I_{n_{2}, n_{1}}(\rho),\quad n_{1} = m,\quad n_{2} = m-l_{\sigma}-\mu,\quad l \leq -(1-\vartheta)/2, \nonumber\\[0.7pt]
\phi_{n_{1}, n_{2}, \sigma}^{(1)}(\theta, \rho) &=& e^{i \left(l_{\sigma} - l_{0}\right) \theta - i\pi l_{\sigma}}\, I_{n_{1}, n_{2}}(\rho),\quad n_{1} = m+l_{\sigma}+\mu,\quad n_{2} = m,\quad l \geq (1+\vartheta)/2,
\end{eqnarray}
where $l_{\sigma} = l - (1+\sigma)/2$, and $I_{m + \alpha, m}(\rho)$ are given by Eq.~(\ref{eq6}). These functions form an orthogonal set on the semiaxes $\rho > 0$ with respect to the scalar product $(f, g)_{\perp}$. The existence of two self-adjoint extensions is correlated with the irregular behavior of the radial functions $I_{n, m}(\rho)$ at the origin when $l = 0$ and either $\sigma = -1$ for $\vartheta = +1$ or $\sigma = +1$ for $\vartheta = -1$ \cite{SPGavrilov03}. Note that $\hat{\Pi}_{0}^{2}\, u_{n_{1}, n_{2},\pm 1}^{(j)} = \mathcal{E}_{\pm}^{2}\, u_{n_{1}, n_{2}, \pm 1}^{(j)}$ then $\hat{\Pi}_{0} \left(  M\right)\, u_{n_{1}, n_{2}, \pm 1}^{(j)} = \mathcal{E}_{\pm}\, u_{n_{1}, n_{2}, \pm 1}^{(j)}$. Thus, the spectrum of the operator $\hat{\Pi}_{0} \left(M\right)  $ is positive defined. We use the $\hat{\Pi}_{0} \left(M\right)$ in  representation (\ref{eq29}) to simplify transition from these stationary states to CS in the next subsection.

With respect to the self-adjoint operators $\hat{H}^{\vartheta}$, it is know that for any $\vartheta$ the functions $\psi_{n_{1}, n_{2}}^{(j)}(\theta, \rho)$
form a complete set on the $xy$-plane.

To prove this directly and to find an explicit form of the unity resolution, we, similar to the non-relativistic case, introduce the retarded Green function $S^{\mathrm{ret}}(x, x^{\prime})$ of the Dirac equation,
\begin{eqnarray}\label{eq32}
S^{\mathrm{ret}}(x, x^{\prime}) &=& \Theta(\Delta t)\left[S^{-}(x, x^{\prime}) + S^{+}(x, x^{\prime})\right] =\, \Theta(\Delta t)[S^{c}(x, x^{\prime}) - S^{\bar{c}}(x, x^{\prime})],\nonumber\\[0.7pt]
S^{c}(x, x^{\prime}) &=& \Theta(\Delta t) S^{-}(x, x^{\prime}) - \Theta(-\Delta
t) S^{+}(x, x^{\prime}),\quad S^{\bar{c}} = \Theta(-\Delta t) S^{-}(x, x^{\prime
}) - \Theta(\Delta t) S^{+}(x, x^{\prime}), \nonumber\\[0.7pt]
S^{\mp}(x, x^{\prime}) &=& i \sum_{j, l, m}\, e^{\mp i\mathcal{E}_{\pm}\Delta
t/\hbar}\, \psi_{\pm, n_{1}, n_{2}}^{(j)}(\theta, \rho)\, \psi_{\pm, n_{1}, n_{2}}^{(j) \dagger}(\theta^{\prime}, \rho^{\prime}) \sigma^{3},
\end{eqnarray}
where summation in Eq.~(\ref{eq32}) is over all possible quantum numbers $j$, $m,$ and $l$ is specified in Eq.~(\ref{eq31}). The functions $S^{c}(x, x^{\prime})$ and $S^{\bar{c}}(x, x^{\prime})$ are the causal and anticausal Green functions, respectively. The resolution of unity is satisfied if in the following relation holds
\begin{equation}\label{eq33}
-i \sigma^{3} \left. S^{\mathrm{ret}}(x, x^{\prime}) \right\vert_{\Delta t=0^{+}} = \delta \left(x - x^{\prime}\right)\, \delta\left(y - y^{\prime}\right)\, \mathbb{I}, 
\end{equation}
where $\mathbb{I}$ is an $2\times 2$ identity matrix. We are going to prove that Eq.~(\ref{eq33}) take place in our case. To this end, we represent $S^{c}(x, x^{\prime})$ and $S^{\bar{c}}(x, x^{\prime})$ in the form of the Fock-Schwinger proper time integral \cite{SPGavrilov04-2}:
\begin{eqnarray}\label{eq34}
S^{c}(x, x^{\prime}) &=& \left[\sigma^{3}(\hat{p}_{0} - \mbox{\boldmath$\sigma$\unboldmath} \mathbf{\hat{P}}_{\perp}) + M c\right]\, \Delta^{c}(x, x^{\prime}), \nonumber\\[0.7pt]
S^{\bar{c}}(x, x^{\prime}) &=& \left[\sigma^{3}(\hat{p}_{0} - \mbox{\boldmath$\sigma$\unboldmath} \mathbf{\hat{P}}_{\perp}) + M c\right]\, \Delta^{\bar{c}}(x, x^{\prime}), \nonumber\\[0.7pt]
\Delta^{c}(x, x^{\prime}) &=& \int_{0}^{\infty} ds\, f(x, x^{\prime
}, s), \quad \Delta^{\bar{c}}(x, x^{\prime}) = \int_{-0}^{-\infty} ds\, f(x, x^{\prime}, s), 
\end{eqnarray}
where $\hat{p}_{0} = \frac{i \hbar}{c} \frac{\partial}{\partial t}$ and the kernel $f(x, x^{\prime}, s)$ is given by
\begin{eqnarray}\label{eq35}
f\left(x, x^{\prime}, s\right) &=& \sum_{\sigma = \pm 1}\, \sum_{l = -\infty}^{\infty} f_{\sigma, l} \left(x, x^{\prime}, s\right), \quad f_{\sigma, l}\left(x, x^{\prime}, s\right) = A_{\sigma, l}\left(s\right)\, B_{\sigma, l}\left(s\right)\, \Xi_{\sigma}, \nonumber\\[0.7pt]
A_{\sigma, l}\left(s\right) &=& \frac{\gamma}{8 \pi^{3/2} s^{1/2} \sin\left(\gamma s\right)} \exp\left\{\frac{i\pi}{4} - i\left(M c \hbar^{-1} \right)^{2} s + i \left(l_{\sigma}-l_{0}\right) \Delta\theta - i \left(l_{\sigma
} + \sigma + \mu \right) \gamma s\right\}  \nonumber\\[0.7pt]
&\times& \exp\left\{-\frac{i \left(c\Delta t\right)^{2}}{4s} + \frac{i}{2} \left(\rho + \rho^{\prime}\right)\cot\left(\gamma s\right)\right\}
\,, \nonumber\\[0.7pt]
B_{\sigma, l}\left(s\right) &=& I_{\left \vert l_{\sigma} + \mu \right \vert} \left(z\right)\; \mathrm{if} \; l\neq 0, \quad B_{\sigma, 0}\left(s\right) = I_{\frac{1+\sigma}{2} - \mu}\left(z\right)\; \mathrm{if}\; \vartheta = +1, \nonumber\\[0.7pt]
B_{\sigma, 0}\left(s\right) &=& I_{\mu - \frac{1+\sigma}{2}}\left(z\right)\; \mathrm{if}\; \vartheta = -1, \quad z = e^{-i\pi/2}\sqrt{\rho \rho^{\prime}}%
/\sin\left(\gamma s\right), \quad \Xi_{\pm 1} = \left(1 \pm \sigma^{3}\right)/2 \,.
\end{eqnarray}
The integration path over $s$ is deformed so that it goes slightly below the singular points $s_{k} = k\pi/\gamma$, and $-s_{k}$, $k = 1, 2, \ldots\, $. Negative values for $s$ are defined as $s = |s|\, e^{-i\pi}$. The kernel $f(x, x^{\prime}, s)$ satisfies the following differential equation
\begin{equation}\label{eq36}
i \frac{d}{ds} f(x, x^{\prime}, s) = \hbar^{-2} \left\{\left(M c \right)^{2} - \left[\sigma^{3} \left(\hat{p}_{0} - \mbox{\boldmath$\sigma$\unboldmath} \mathbf{\hat{P}}_{\perp}\right) \right]^{2}\right\} \, f(x, x^{\prime}, s).
\end{equation}

Remembering that $S^{\mathrm{ret}}(x, x^{\prime})$ is the integral kernel of an integral over the variables $\rho$, $\rho^{\prime}$, we can fulfill a transformation used above in the non-relativistic case and change the integral path on the complex plane of $\rho$, $\rho^{\prime}$ to a form where $\rho  = i\xi$, $\rho^{\prime} = i \xi^{\prime}$ with real positive $\xi$, $\xi^{\prime}$. With respect to irregular behavior of the quantities $B_{\sigma, 0}\left(s\right)$ as $\left\vert \rho \rho^{\prime}\right\vert \rightarrow 0$, we restrict the range of $\left\vert \rho \rho^{\prime}\right\vert$ to $0 < \delta < \left\vert \rho \rho^{\prime} \right\vert < \infty$, with arbitrary $\delta\ll 1$. Then we take the limit $s\rightarrow 0^{+}$. Under such a condition we use the asymptotic expansion of the Bessel function as $s\rightarrow 0^{+}$ and then return to the original variable $\rho$, $\rho^{\prime}$ on the real semiaxes. Thus, we find that
\begin{eqnarray}\nonumber
\lim_{s\rightarrow 0^{+}} f_{\sigma, l}\left(x, x^{\prime}, s\right) = \frac{i c \hbar \gamma}{2 \pi} e^{i \left(l_{\sigma} - l_{0}\right) \Delta \theta} \delta \left(c\Delta t\right) \delta(\rho - \rho^{\prime})\, \Xi_{\sigma}\nonumber
\end{eqnarray}
for both self-adjoint extensions $\vartheta = \pm 1$. Finally, using representation (\ref{eq13}), we obtain
\begin{equation}\label{eq37}
\lim_{s\rightarrow 0^{+}} f(x, x^{\prime}, s) = i c \hbar\, \delta\left(c\Delta
t\right)\, \delta(x - x^{\prime})\, \delta(y - y^{\prime}) \mathbb{I}. 
\end{equation}
In the same manner, we can get
\begin{equation}\label{eq38}
\lim_{s\rightarrow 0^{-}}f(x, x^{\prime}, s) = -i c \hbar\, \delta\left(c\Delta
t\right)\, \delta(x - x^{\prime})\, \delta(y - y^{\prime})\mathbb{I}. 
\end{equation}

Taking into account that the kernel $f(x, x^{\prime}, s)$ has no any singularity
in the lower part of complex plane of $s$, the integral $\Delta^{\mathrm{ret}}(x, x^{\prime}) = \Theta(\Delta t)\left[\Delta^{c}(x, x^{\prime}) - \Delta^{\bar{c}}(x, x^{\prime})\right]$ can be represented as
\begin{equation}\label{eq39}
\Delta^{\mathrm{ret}}(x, x^{\prime}) = \Theta(\Delta t) \int_{\Gamma} ds\, f(x, x^{\prime}, s), 
\end{equation}
where $\Gamma$ is a clockwise circle, which connects the points $s = +0$ and $s = 0\cdot e^{-i\pi}$, and passes in the lower part of complex plane of $s$. If conditions (\ref{eq37}) and (\ref{eq38}) hold, the function
\begin{equation}\label{eq40}
S^{\mathrm{ret}}(x, x^{\prime}) = \left[\sigma^{3}(\hat{p}_{0} - \mbox{\boldmath$\sigma$\unboldmath} \mathbf{\hat{P}}_{\perp}) + M c\right] \, \Delta^{\mathrm{ret}}(x, x^{\prime})
\end{equation}
satisfies Eq. (\ref{eq33}) and is indeed the retarded Green function of the corresponding Dirac equation \cite{SPGavrilov96}.

\section{Relativistic CS}

Generally speaking, in the relativistic case, the Dirac Hamiltonian is not quadratic in the momenta. Due to this fact the time evolution of instantaneous CS on $xy$-plane (see details in \cite{VGBagrov11}) is not trivial. However, because the time evolution of these states is unitary, it is enough to show that the set of such initial CS is complete.

For instance, CS for massive spinning ('spin up') particle in MSF on the $xy$-plane and in ($2+1$) dimension are
\begin{equation}\label{eq41}
\mathbf{\Psi}_{\pm, z_{1}, z_{2}}^{(j)}(\theta, \rho) = \left\{\sigma^{3} \left[
\pm \hat{\Pi}_{0} \left(M\right) - \mbox{\boldmath$\sigma$\unboldmath} \mathbf{\hat{P}}_{\perp}\right] + Mc\right\} \mathbf{u}_{z_{1}, z_{2}, \pm 1}^{(j)}(\theta, \rho),\quad \mathbf{u}_{z_{1}, z_{2}, \sigma}^{(j)}(\theta, \rho) = \mathbf{\Phi}_{z_{1}, z_{2}, \sigma}^{(j)}(\theta, \rho)\,v_{\sigma}, 
\end{equation}
where $\mathbf{\Phi}_{z_{1}, z_{2}, \sigma}^{(j)}(\theta, \rho)$ are defined in the similar way as the CS of non-relativistic electron, see Eqs.~(\ref{eq14}). Taking Eq.~(\ref{eq29}) into account, one can see that the CS (\ref{eq41}) can be written as
\begin{eqnarray}\label{eq42}
\mathbf{\Psi}_{\pm, z_{1}, z_{2}}^{(j)}(\theta, \rho) &=& \frac{1}{\sqrt{\mathcal{M}_{j, \pm}(|z_{1}|^{2}, |z_{2}|^{2})}} \sum_{l} \mathbf{\Psi}_{\pm, z_{1}, z_{2}}^{(j), l}(\theta, \rho), \nonumber\\
\mathbf{\Psi}_{\pm, z_{1}, z_{2}}^{(j), l}(\theta, \rho) &=& \sum_{m = 0}^{\infty} \frac{z_{1}^{n_{1}}\, z_{2}^{n_{2}}}{\sqrt{\Gamma(1 + n_{1}) \Gamma(1+n_{2})}}\, \psi_{\pm, n_{1}, n_{2}}^{(j)}(\theta, \rho),
\end{eqnarray}
where $n_{1}$, $n_{2}$, $j$ and $l$ change according to Eq.~(\ref{eq31}). The normalization constants $\mathcal{M}_{j, \pm}(|z_{1}|^{2}, |z_{2}|^{2})$ can be calculated from the overlapping formula
\begin{eqnarray}\nonumber
\left(\mathbf{\Psi}_{\pm, z_{1}, z_{2}}^{(j)}, \mathbf{\Psi}_{\pm, z`_{1}, z`_{2}}^{(j)}\right)_{D} = \frac{2 M c}{\sqrt{\mathcal{M}_{j, \pm}(|z_{1}|^{2}, |z_{2}|^{2}) \mathcal{M}_{j, \pm}(|z_{1}^{\prime}|^{2}, |z_{2}^{\prime}|^{2})}} \left(\mathbf{\Phi}_{z_{1}, z_{2},\pm 1}^{(j)}, \left[\pm \hat{\Pi}_{0} \left( M\right) + M c\right] \mathbf{\Phi}_{z_{1}^{\prime}, z_{2}^{\prime}, \pm 1}^{(j^{\prime})}\right)_{\bot},
\end{eqnarray}
where the inner products $\left(\cdot, \cdot\right)_{D}$ and $\left(\cdot, \cdot\right)_{\bot}$ are defined by (\ref{eq30}) and (\ref{eq5}), respectively.

Representation (\ref{eq42}) is like (\ref{eq14}), so that the unity resolution in CS can be done in a form similar to Eq.~(\ref{eq17}). Taking into account the structure of the retarded Green function $S^{\mathrm{ret}}(x, x^{\prime})$ given by (\ref{eq32}), we find
\begin{eqnarray}\label{eq43}
& & \sum_{j = 0, 1} \left.F^{(j)}(x, x^{\prime}) \right\vert_{\Delta t=0^{+}} = -i \sigma^{3}\left. S^{\mathrm{ret}}(x, x^{\prime}) \right\vert_{\Delta t=0^{+}}, \quad k = 1, 2, \nonumber\\
& & F^{(j)}(x ,x^{\prime}) = \int d^{2}z_{1}\, d^{2}z_{2}\, W_{j}^{\mu}(|z_{1}|^{2}, |z_{2}|^{2}) \sum_{\zeta = \pm}\, e^{-i \zeta c \hat{\Pi}_{0}\left(M\right) \Delta t/\hbar} \mathbf{\Psi}_{\zeta, z_{1}, z_{2}}^{(j)}(\theta, \rho)\mathbf{\Psi}_{\zeta, z_{1}, z_{2}}^{(j)\dagger}(\theta^{\prime}, \rho^{\prime}),
\end{eqnarray}
where the weight function $W_{j}^{\mu}(|z_{1}|^{2}, |z_{2}|^{2})$ is the same as in (\ref{eq17}). As in the non-relativistic case, the consistent limit as $\Delta t\rightarrow 0^{+}$ can be considered due to the inclusion of an appropriate time-dependent phase factor in the definition of $F^{(j)}(x, x^{\prime})$ in (\ref{eq43}). In the relativistic case under consideration, we have two such different factors, one for particle states, another one for antiparticle states. The proof of the resolution identity (\ref{eq43}) is quite similar to the one for $\Phi_{z_{1}, z_{2}}^{(j)}(\theta, \rho)$. Using the same weight function $W_{j}^{\mu}(|z_{1}|^{2}, |z_{2}|^{2}),$ we obtain a similar result for the massless fermions.

To complete our consideration, we consider $(3+1)$ case. The domains of $(3+1)$-Dirac Hamiltonian in MSF are trivial extensions of the corresponding  domains mentioned in $(2+1)$ case, that is why we use for self-adjoint $(3+1)$-Dirac Hamiltonian the same notation $\hat{H}^{\vartheta}$. Of course, in this case $\hat{H}^{\vartheta} = c\gamma^{0} \left(\sum_{k = 1, 2, 3} \gamma^{k} \hat{P}^{k} + M c\right)$, $\,\hat{P}^{3} = i \hbar \partial_{z}$, and $\gamma^{0}$, $\gamma^{k}$ are the $4\times 4$ Dirac gamma matrices. In particular, we consider the CS for spinning particle in $(3+1)$-dim, which can be constructed by using the set of the orthogonal stationary states defined in Eq.~(A.39) from \cite{VGBagrov11}. The latter can be reduced to ones in $(2+1)$-dim as follows
\begin{eqnarray}\label{eq44}
\Psi_{\pm, s, p_{3}, n_{1}, n_{2}}^{(j)}(x) &=& \exp\left[-\frac{i}{\hbar}(c \hat{\Pi}_{0} \left(\tilde{M}\right) t + p_{3} z) \right] \Psi_{\pm, s, p_{3}, n_{1}, n_{2}}^{(j)}(x_{\perp})\,,\quad s = \pm 1,\nonumber\\[0.25in]
\Psi_{\pm, s, p_{3}, n_{1}, n_{2}}^{(j)}(x_{\perp}) &=& \mathcal{M}_{p_{3}} \left(
\begin{array}[c]{l}
\left[M^{-1}\left(p^{3}/c + s\widetilde{M}\right) + 1\right]\tilde{\psi}_{\pm, n_{1}, n_{2}}^{(j)}(\theta, \rho)\\
\mbox{} \\
\left[M^{-1}\left(p^{3}/c + s\widetilde{M}\right) - 1\right]\sigma^{3} \tilde{\psi}_{\mp, n_{1}, n_{2}}^{(j)}(\theta, \rho)
\end{array}
\right)\,.
\end{eqnarray}
Here $\widetilde{M} = \sqrt{M^{2} + (p_{3}/c)^{2}}$; $p_{3}$ is $z$-component of
the momentum; $\hat{\Pi}_{0} (\widetilde{M}) = \left.\hat{\Pi}_{0}(M)\right\vert _{M = \widetilde{M}}$, where $\hat{\Pi}_{0}(M)$ is defined in (\ref{eq29}); $\mathcal{M}_{p_{3}}$ is an additional normalization factor; $s$ are eigenvalues of the spin operator $\hat{S}_{z}$,
\begin{eqnarray}\nonumber
\hat{S}_{z} = \frac{1}{2} \left(\hat{H}^{\vartheta} \Sigma_{z} + \Sigma_{z} \hat{H}^{\vartheta} \right)/\widetilde{M}c^{2};
\end{eqnarray}
and $\tilde{\psi}_{\pm, n_{1}, n_{2}}^{(j)}(\theta, \rho) = \left. \psi_{\pm, n_{1}, n_{2}}^{(j)}(\theta, \rho) \right\vert_{M = \widetilde{M}}$, where $\psi_{\pm, n_{1}, n_{2}}^{(j)}(\theta, \rho)$ is defined in (\ref{eq29}).

Using the matrix structure (\ref{eq44}) we can find instantaneous CS in $(3+1)$-dim in the similar form
\begin{equation}\label{eq45}
\mathbf{\Psi}_{\pm, s, p_{3}, z_{1}, z_{2}}^{(j)}(\theta, \rho)  = \mathcal{M}_{p_{3}} \left(
\begin{array}[c]{l}
\left[M^{-1}\left(p^{3}/c + s\widetilde{M}\right) + 1\right] \widetilde{\mathbf{\Psi}}_{\pm, z_{1}, z_{2}}^{(j)}(\theta, \rho)\\
\mbox{} \\
\left[M^{-1}\left(p^{3}/c + s\widetilde{M}\right) - 1\right] \sigma^{3} \widetilde{\mathbf{\Psi}}_{\mp, z_{1}, z_{2}}^{(j)}(\theta, \rho)
\end{array}
\right)\,.
\end{equation}
Here the two component column $\widetilde{\mathbf{\Psi}}_{\pm, z_{1}, z_{2}}^{(j)}(\theta, \rho)$ is defined as
\begin{eqnarray}\nonumber
\widetilde{\mathbf{\Psi}}_{\pm, z_{1}, z_{2}}^{(j)}(\theta, \rho) = \left. \mathbf{\Psi}_{\pm, z_{1}, z_{2}}^{(j)}(\theta, \rho) \right\vert_{M = \widetilde{M}},
\end{eqnarray}
where $\mathbf{\Psi}_{\pm, z_{1}, z_{2}}^{(j)}(\theta, \rho)$ are CS in $(2+1)$-dim given by (\ref{eq42}). Thus, we see that for given $p_{3}$ and $s$ the unity resolution in CS for $(3+1)$-dim case is reduced to $(2+1)$-dim considered above. Note that representations (\ref{eq44}) and (\ref{eq45}) are convenient for the non-relativistic limit.

Another types of instantaneous CS in $(3+1)$-dim that allow one to construct relativistic time-dependent CS were obtained in the work \cite{VGBagrov11}, see Eq.~(40) there. They have another matrix structure. In the same work it was demonstrated that instantaneous CS in $(3+1)$-dim are reduced to ones in $(2+1)$-dim. That is why the unity resolution in terms of CS in $(3+1)$-dim is also reduced to $(2+1)$-dim case considered above. This allows one to prove that the relativistic time-dependent CS given by Eq.~(89) from \cite{VGBagrov11} form a complete system on the light cone hypersurface $ct - z = \mathrm{const}$.

\subparagraph{{\protect\large Acknowledgement}}

KG acknowledges support from Funda\c{c}\~{a}o de Amparo \'{a} Pesquisa do
Estado de S\~{a}o Paulo (FAPESP, Brazil) under Program No.~2010/15698-5; VGB
thanks FAPESP (Brazil) and Russian Science and Innovations Federal Agency
under contract No 02.740.11.0238 and Russia President grant SS-3400.2010.2 for
support; SPG acknowledges support of the program Bolsista CAPES/Brazil and
thanks the University of S\~{a}o Paulo for hospitality. DMG acknowledges the
permanent support of FAPESP and CNPq.

\end{document}